\documentclass[twocolumn,showpacs,preprintnumbers,amsmath,amssymb]{revtex4}
\usepackage[dvips]{graphicx}
\usepackage{dcolumn}

\begin{document}

\title
{Orbital-limiting and modulation induced by missing parity in non-centrosymmetric superconductors}

\author{Norihito Hiasa and Ryusuke Ikeda}

\affiliation{%
Department of Physics, Kyoto University, Kyoto 606-8502, Japan
}

\date{\today}

\begin{abstract} 
We examine the depairing field $H_{c2}(T)$ of non-centrosymmetric superconductors with a spin-orbit coupling larger than the Zeeman energy at $H_{c2}(0)$ by taking account of the mixing of spin-singlet and triplet pairing states due to the missing parity. When the singlet and triplet pairing components are mixed with an equal weight in a cubic non-centrosymmetric system, the paramagnetic depairing effect is significantly suppressed so that $H_{c2}(T)$ approaches its orbital-limited value. A similar event also occurs in a quasi two-dimensional Rashba non-centrosymmetric system. The present results are relevant to the $H$-$T$ phase diagrams of CePt$_3$Si and the families of Li$_2$Pd$_{3-x}$Pt$_x$B. 
\end{abstract}

\pacs{}

%\keyword{}

\maketitle

\section{Introduction}

A Cooper-pair condensate in a superconductor is destabilized by two kinds of field-induced mechanisms, the paramagnetic depairing and the orbital depairing. The orbital-limited case, i.e., a situation with a depairing field $H_{c2}(T)$ determined only by the increase of the number of vortices, conventionally occurs for spin-triplet superconductors with an equal-spin pairing. Recently, it has been noticed that the orbital-limited case occurs in Rashba non-centrosymmetric superconductors \cite{Frigeri} in an applied field (${\bf H} \parallel c$) perpendicular to the basal plane on the space inversion asymmetry \cite{Kaur}. In contrast, the paramagnetic depairing is so effective in Rashba superconductors in the parallel field ${\bf H} \perp c$ that, as seen in CeRhSi$_3$ \cite{Kimura} and CeIrSi$_3$ \cite{Settai}, $H_{c2}(T)$ is strongly suppressed, and that the vortex state shows peculiar modulated structures \cite{ymatsu}. However, it should be noted that these results in the parallel field were obtained in the case with a purely singlet (or purely triplet) pairing. Originally, a hallmark of non-centrosymmetric superconductors is a missing parity and the resulting coexistence of spin-singlet and triplet pairing symmetries \cite{Frigeri}. However, this mixing of pairing symmetries and a coupling between them have not been taken into account so far in studying $H_{c2}(T)$ and the vortex states occurring below it. In relation to this, it should be noted that the strong anisotropy of $H_{c2}$ due to the anisotropic paramagnetic effect does not seem to be a common feature between Rashba superconductors: A strongly suppressed $H_{c2}$ in the parallel field case was realized in a couple of materials \cite{Kimura,Settai} with tetragonal structure, while the $H_{c2}$-curves in CePt$_3$Si are nearly isotropic \cite{Yasuda}.  

In this work, we study the depairing field $H_{c2}(T)$ of non-centrosymmetric 
superconductors with coexisting singlet and triplet pairing components. Both of systems with the spin-orbit coupling of Rashba type in ${\bf H} \perp c$ and those with that of cubic type will be considered here because a comparison between those two cases is found to be useful. Following previous works \cite{Kaur,ymatsu,Samokhin}, a moderately large value of the spin-orbit coupling $\zeta$ will be assumed, Max($T$, $\mu_{\rm B} H$) $\ll \zeta \ll E_{\rm F}$, where $\mu_{\rm B} H$ is the Zeeman energy of a conduction electron, and $E_{\rm F}$ is the averaged Fermi energy. In such materials in a single pairing channel, a one-dimensional modulation \cite{Kaur,ymatsu} of the superconducting order parameter $\Delta$ and the resulting increase of $H_{c2}$ occur as a result of the small but nonvanishing $\zeta/E_{\rm F}$. We show below that a coupling {\it induced by spatial variations of} $\Delta$ between the coexisting singlet and triplet pairing channels is a much stronger origin of elevating $H_{c2}$-values. When the two pairing channels are equally important, one of the two FSs splitted due to the missing parity is favored for the gap formation, and this imbalance between the FSs leads to a more perfect disappearance of the paramagnetic depairing than that due to a finite $\zeta/E_{\rm F}$ \cite{Kaur,ymatsu}. In fact, under proper conditions, the present mechanism based on the missing parity results even in the orbital-limited situation where the paramagnetic depairing is quenched. It is argued that the present results should be closely related to the fact that the $H_{c2}$-anisotropy in Rashba superconductors significantly depends on the materials \cite{Kimura,Settai,Yasuda} and are also relevant to the $x$-dependence of pairing states of Li$_2$(Pd$_{3-x}$Pt$_x$)B \cite{Yuan}. 

Through this paper, the $H_{c2}$-enhancement due to the mixing of the singlet and triplet pairing channels is discussed, for clarity, by focusing primarily on the cubic case. In the cubic case, the $H_{c2}$-enhancement is accompanied, as in the centrosymmetric case with the Fulde-Ferrell (FF) state \cite{FF}, by a helical modulation parallel to ${\bf H}$ of the phase of $\Delta$. It will be pointed out that this cubic case corresponds to an ideal situation of the familiar FFLO mechanism of an $H_{c2}$-enhancement in which the paramagnetic depairing is cancelled by a modulation of $\Delta$. 
In addition, it is pointed out that, in this cubic case, the Larkin-Ovchinnikov (LO) state with a periodic amplitude modulation \cite{LO} parallel to ${\bf H}$ cannot obtain a gain in energy necessary for its realization. 

This paper is organized as follows. In Sec.II, the starting model and the formulation are explained, and possible $H_{c2}(T)$-lines in the cubic noncentrosymmetric case are discussed. As an observable quantity measuring the broken inversion symmetry in the cubic case, a transverse component of the local magnetization in the vortex lattice is discussed in sec.III. In sec.IV, the $H_{c2}(T)$-curves in the Rashba superconductors under a field {\it parallel} to the basal plane are considered for a comparison with the results in sec.II. A summary and some comments are given in sec.V. 

% $T^{(0)}_s$ ($T^{(0)}_t$) is the zero field SC transition temperature of the s%inglet (triplet) pairing component in the absence of the triplet (singlet) comp%onent. 

\section{Model and cubic case}
 
We start from the following electronic Hamiltonian 
\begin{eqnarray}
{\cal H}_{el} &=& \sum_{{\bf k}, s_1, s_2} c_{{\bf k}, s_1}^\dagger (  \varepsilon_{\bf k} \, \delta_{s_1, s_2} + (\zeta \, {\hat {\bf g}}_{\bf k} + \mu_{\rm B} {\bf H})\cdot{\bf \sigma}_{s_1, s_2} \, ) \, c_{{\bf k}, s_2} \nonumber \\
&+& \frac{1}{V} \! \sum_{{\bf p}, {\bf k}_1, {\bf k}_2} \! W_{\alpha \, \beta, \, \gamma \, \delta}({\bf k}_1, {\bf k}_2) \, c^\dagger_{{\bf k}_1+{\bf p}/2, \alpha} \, c^\dagger_{-{\bf k}_1+{\bf p}/2, \beta} \nonumber \\
&\times& c_{-{\bf k}_2+{\bf p}/2, \delta} \, c_{{\bf k}_2+{\bf p}/2, \gamma},
\end{eqnarray}
where $\mu_{\rm B} H$ is the Zeeman energy, $\varepsilon_{\bf k}$ is the bare band energy, ${\bf \sigma}_{\alpha, \beta}$ are the Pauli matrices, $V$ is the volume, and ${\hat {\bf g}}_{\bf k}$ parametrizes the spin-orbit coupling. The gauge field will be incorporated later at the quasi-classical level. The pairing interaction is represented by 
\begin{equation}
W_{\alpha \, \beta, \, \gamma \, \delta}({\bf k}_1, {\bf k}_2) = - \frac{1}{2} \sum_{i,j=s,t} w_{ij} ({\hat \tau}^\dagger_i({\bf k}_1))_{\alpha \, \beta} \, ({\hat \tau}_j({\bf k}_2))_{\delta \, \gamma},
\end{equation}
where $\tau_s({\bf k})={\rm i} \sigma_y$, $\tau_t({\bf k}) = {\rm i} (\sigma_y \sigma_\mu)\cdot({\hat {\bf g}}_{\bf k})_\mu$, the $2 \times 2$ matrix $w_{ij}$ is positive definite, and the index $s$ ($t$) implies the spin-singlet (triplet) component. It has been assumed in eq.(2) that $|\zeta|$ is so large that the spin component, i.e., $d$-vector, of $\tau_t({\bf k})$ is protected by the spin-orbit coupling \cite{Frigeri}. By diagonalizing the quadratic term of ${\cal H}_{el}$ through the unitary transformation $c_{{\bf k},\beta} = \sum_{a=1,2} U_{\beta a}({\bf k}) d_{{\bf k},a}$, the single particle energy close to the Fermi surface (FS) $a$ ($=1$, $2$) is given by $\varepsilon_{\bf k} - (-)^a \zeta|{\hat {\bf g}}_{\bf k}|$, and the interaction Hamiltonian (the second line of eq.(1)) consistently takes the form 
\begin{equation}
{\cal H}_{\rm int} = -\frac{V}{2} \sum_{\bf p} \sum_{i,j=s,t} w_{ij} [\Psi^{(i)}_{\bf p}]^\dagger \Psi^{(j)}_{\bf p}. 
\end{equation}

In the cubic case, ${\hat {\bf g}}_{\bf k}$ simply becomes ${\hat k}={\bf k}/k_F = {\hat z} {\rm cos}\theta_{\bf k} + {\rm sin}\theta_{\bf k} ({\hat x} {\rm cos}\phi_{\bf k} + {\hat y} {\rm sin}\phi_{\bf k})$ up to the lowest order in ${\bf k}$, and, following other works \cite{Yip}, this model will be used here. The transformation matrix $U({\bf k})$ takes the form ${\rm cos}(\theta_{\bf k}/2) + {\rm i} \, {\rm sin}(\theta_{\bf k}/2) ({\rm sin}\phi_{\bf k} \sigma_x - {\rm cos}\phi_{\bf k} \sigma_y)$. Then, $\Psi^{(j)}_{\bf p}$ is expressed by  
\begin{eqnarray}
\Psi^{(t)}_{\bf p} &=&  \sum_{\bf k} \frac{|{\hat {\bf g}}_{\bf k}|}{V} \sum_{a} e^{{\rm i}(\pi (a+1) + (-1)^{a+1} \phi_k)} \, d_{-{\bf k}+{\bf p}/2, a} 
d_{{\bf k}+{\bf p}/2, a} \nonumber \\
\Psi^{(s)}_{\bf p} &=& - \frac{{\rm i}}{V} \sum_{\bf k} \sum_a e^{{\rm i}(-1)^{a+1} \phi_k} \, d_{-{\bf k}+{\bf p}/2, a} d_{{\bf k}+{\bf p}/2, a}. 
\end{eqnarray}
In eq.(3), O($\varepsilon_H$) corrections expressing an interband pairing were neglected in writing ${\cal H}_{\rm int}$, where $\varepsilon_H = {\rm Max}(\mu_{\rm B} H, \,\, T)/|\zeta|$. In the Ginzburg-Landau  (GL) free energy $F^{(c)}$ given below, they would lead to a correction term of O($\varepsilon_H^3$) which is safely negligible. 
Then, by decoupling ${\cal H}_{\rm int}$ in the manner $- \sum_{i,j}w_{ij} [\Psi^{(i)}]^\dagger \Psi^{(j)} \to \sum_{i,j} [ (w^{-1})_{ij} \Delta_i^* \Delta_j ] - (\Delta_s^* \Psi^{(s)} + \Delta_t^* \Psi^{(t)} + {\rm h.c.})$, the resulting $F^{(c)}$ in the cubic case can be represented as a functional of the order parameters $\Delta_a$ ($a=1$, $2$) defined on 
the resulting two FSs, where  
\begin{equation}
\Delta_a = \frac{- {\rm i} \Delta_s + (-1)^a \Delta_t}{\sqrt{2}}.  
\end{equation}
In obtaining $F^{(c)}$, one needs to use the expression 
\begin{equation}
{\cal G}_a({\bf k}, {\rm i}\varepsilon) = \frac{1}{{\rm i}\varepsilon - \varepsilon_{\bf k} + (-1)^a|\zeta {\hat {\bf g}}_{\bf k} + \mu_{\rm B} {\bf H}|}.
\end{equation}
Then, using the relation 
\begin{eqnarray}
&\int&\frac{d^3{\bf k}}{(2 \pi)^3} \, {\cal G}_a({\bf k}, {\rm i}\varepsilon) {\cal G}_a(-{\bf k}+{\bf \Pi}, -{\rm i}\varepsilon) \nonumber \\
&=& \frac{2 \pi N_a}{2|\varepsilon| + {\rm i} {\rm sgn}\varepsilon ({\bf v}\cdot{\bf \Pi} + 2 (-1)^{a+1} \mu_{\rm B} {\bf H}\cdot{\hat {\bf g}}_{\bf k})}, 
\end{eqnarray}
where $N_a$ is the density of states on the FS $a$, the quadratic term of the GL free energy $F^{(c)}$ is given by 
\begin{eqnarray}
F^{(c)}_2 \! &=& \! \int d^3r \biggl[ \, \sum_{a=1,2} [ \, ((w^{-1})_{ss} + (w^{-1})_{tt}) |\Delta_a|^2 - 2 \Delta^*_a K_a({\bf \Pi}) \Delta_a \,] \nonumber \\
&+& [ ( \, (w^{-1})_{ss} - (w^{-1})_{tt} - 2 {\rm i} \, (w^{-1})_{st} \, ) \Delta_1^* \Delta_2 + {\rm c.c.} ] \, \biggr],
\end{eqnarray}
where 
\begin{eqnarray}
K_a({\bf \Pi}) = 2 N_a \int_{\rho_c}^\infty d\rho \, f(\rho; T) \langle {\rm cos}(\rho {\bf v}\cdot{\bf \Pi}_a) \rangle,
\end{eqnarray}
\begin{equation}
f(\rho; T) = \frac{2 \pi T}{{\rm sinh}(2 \pi 
T \rho)}, 
\end{equation}
${\bf \Pi}_a = {\bf \Pi} - (-)^a Q {\hat z}$ in a field ${\bf H} \parallel {\hat z}$, ${\bf \Pi}= - {\rm i}\nabla + 2e {\bf A}$, $Q=2 \mu_{\rm B} H/|{\bf v}|$, ${\bf v}$ is the Fermi velocity vector, $\langle \,\,\, \rangle$ denotes the (angle) average over each FS, and the identity $D^{-1} = \int_0^\infty d\rho \exp(-\rho D)$ was used. In eq.(9), a lower cutoff $\rho_c$ of the $\rho$-integral, which is of the order of the inverse of a high energy cutoff $\omega_c$, was introduced. This will be needed even in some of the ensuing expressions. Note that, in the present case with the cubic spin-orbit coupling breaking the inversion symmetry, the paramagnetic effect appears only through the $Q$-dependence, which simply shifts the gauge field parallel to ${\bf H}$ in a way {\it dependent on} FS (see the sign factor $(-1)^a$ in ${\bf \Pi}_a$). 

To obtain the depairing field $H_{c2}(T)$, we express the order parameter as 
\begin{equation} 
\Delta_a = \frac{1}{\sqrt{L_z}} Z_a({\bf q}) e^{{\rm i}qz} \varphi_0(x,y)
\end{equation}
\cite{FF} and diagonalize $F^{(c)}_2$ with respect to $Z_a$, where $\varphi_0(x,y)$ is an Abrikosov lattice solution in the lowest ($n=0$) Landau level (LL). This restriction to the lowest LL is justified as follows: It is well known that the paramagnetic depairing tends to reverse roles of the lowest LL and higher LLs \cite{AI,I1}. In the present cubic case, however, the paramagnetic effect merely plays the role of modulation parallel to ${\bf H}$ and is ineffective for spatial variations of $\Delta$ perpendicular to ${\bf H}$. In this sense, this situation is similar to the familiar orbital-limited case, and our neglect of higher ($n \geq 1$) LLs in this section is safely valid at least as far as focusing on properties occuring near $H_{c2}$. Then, after diagonalizing $F^{(c)}_2$, the eigenvalue determining the $H_{c2}(T)$-line is given by 
\begin{widetext}
\begin{eqnarray}
\frac{(w^{-1})_{ss} + (w^{-1})_{tt}}{2(N_1+N_2)} &-& \int_{\rho_c}^\infty d\rho \, f(\rho; T) \langle I_0(\rho) ({\rm cos}(\rho v_\parallel q) {\rm cos}(\rho v_\parallel Q) 
+ \delta N {\rm sin}(\rho v_\parallel q) {\rm sin}(\rho v_\parallel Q) ) \rangle \nonumber \\
&=& \biggl[ \delta^2 + \biggl(\int_{\rho_c}^\infty d\rho \, f(\rho; T) \langle I_0(\rho) [ \, \delta N {\rm cos}(\rho v_\parallel q) {\rm cos}(\rho v_\parallel Q) + {\rm sin}(\rho v_\parallel q) {\rm sin}(\rho v_\parallel Q) \, ]  \rangle \biggr)^2 \biggr]^{1/2}, 
\end{eqnarray}
\end{widetext}
where 
\begin{equation}
I_n(\rho) = \exp\biggl(-\frac{\rho^2 |{\bf v}_\perp|^2}{4 r_H^2} \biggr) \, L_n( \rho^2 |{\bf v}_\perp|^2/(2 r_H^2)), 
\end{equation}
$r_H = (2 e H)^{-1/2}$, ${\bf v}_\perp$ ($v_\parallel$) is the component of ${\bf v}$ perpendicular (parallel) to ${\bf H}$, $L_n(x)$ is the Laguerre polynomial, and 
\begin{eqnarray}
(N_1+N_2) \delta = \! \biggl([(w^{-1})_{st}]^2 + \frac{[ (w^{-1})_{ss} - (w^{-1})_{tt} ]^2}{4} \biggr)^{1/2}. 
\end{eqnarray}
Further, by reexpressing eq.(12) in terms of the zero field transition temperature $T_c$ which is determined from eq.(12) in $H=0$ case, we find that the $H_{c2}(T)$-curve is given by  
\begin{widetext}
\begin{eqnarray}
{\rm ln}\biggl(\frac{T}{T_c} \biggr) \! &+& \! \int_0^\infty d\rho \, f(\rho; T) \biggl[\, 1 - \langle I_0(\rho) ({\rm cos}(\rho v_\parallel q) {\rm cos}(\rho v_\parallel Q) + \delta N {\rm sin}(\rho v_\parallel q) {\rm sin}(\rho v_\parallel Q) ) \rangle \biggr] 
= \biggl[ \delta^2 + \biggl(\int_{\rho_c}^\infty d\rho \, f(\rho; T) \nonumber \\ 
&\times& \langle I_0(\rho) [ \, \delta N {\rm cos}(\rho v_\parallel q) {\rm cos}(\rho v_\parallel Q) + {\rm sin}(\rho v_\parallel q) {\rm sin}(\rho v_\parallel Q) \, ]  \rangle \biggr)^2 \biggr]^{1/2} - \biggl[ \delta^2 + \biggl(\int_{\rho_c}^\infty d\rho \, \delta N \, f(\rho; T_c) \biggr)^2 \biggr]^{1/2} 
\end{eqnarray}
\end{widetext}
if $q$ is chosen so that the highest $H$-value results in, where $\delta N= (N_2-N_1)/(N_1+N_2)$. The coefficient $Z_a$ for the corresponding eigenstate is given by 
\begin{equation}
Z_a(q) = \sqrt{( 1 + (-1)^a {\rm sgn}(\delta N) R^{-1/2} )/2},
\end{equation} 
where  
\begin{widetext}
\begin{equation}
 R = 1 + \delta^2 \, \biggl(\int_{\rho_c}^\infty d\rho f(\rho; T) \langle I_0(\rho) [ \, \delta N {\rm cos}(\rho v_\parallel q) {\rm cos}(\rho v_\parallel Q) + {\rm sin}(\rho v_\parallel q) {\rm sin}(\rho v_\parallel Q) \, ]  \rangle \biggr)^{-2}. 
\end{equation}
\end{widetext}

The off-diagonal term $|(w^{-1})_{st}|/(N_1+N_2)$ is a consequence of the lack of inversion symmetry and will be nonzero in general \cite{Frigeri2}. We expect it to be at most of the order $|\delta N| {\rm ln}(\omega_c/T_c)$. 
If {\it both} $(w^{-1})_{st}/(N_1+N_2)$ and $|(w^{-1})_{tt} - (w^{-1})_{ss}|/[2(N_1+N_2)]$ are negligibly small, eq.(15) reduces to its $\delta=0$ case 
\begin{eqnarray}
{\rm ln}\biggl(\frac{T}{T_c} \biggr) &+& \int_0^\infty d\rho f(\rho; T) [ \, 1 - \langle I_0(\rho) \nonumber \\ 
&\times& {\rm cos}(\rho v_\parallel (q + Q)) \rangle \, ] = 0  
\end{eqnarray}
irrespective of $|\delta N|$, where it was assumed that $\delta N < 0$. Then, $H_{c2}(T)$ is given by eq.(18) {\it with} $q+Q=0$, which is the purely orbital-limited one {\it independent} of the paramagnetic depairing. In this specific case, $\Delta_1$ grows on cooling under the condition $\delta N < 0$ ($N_1 > N_2$), while $\Delta_2$ identically vanishes. Further, the period of the {\it phase} modulation is precisely $2 \pi/Q = \pi |{\bf v}|/(\mu_{\rm B} H)$. Since this $Q$ is nothing but the relative shift of the two FSs, this $\delta = 0$ limit in the cubic case can be regarded as an ideal case of the familiar FFLO mechanism for an $H_{c2}$-enhancement: In the present case, the effect of the paramagnetic depairing is perfectly cancelled by the phase modulation (i.e., a nonzero $|q|$) to reach the orbital-limited case. As far as $|(w^{-1})_{st}/(N_1+N_2)|$ will be nonzero, however, the growth of $\Delta_2$ is not negligible even if $w_{ss}=w_{tt}$. 

In place of the helical variation $\Delta_a \sim e^{{\rm i}qz}$, we have also examined the alternative $z$-dependence, $\Delta_a \sim \exp({\rm i}(-)^{a}Qz)$, which is motivated by the fact that the gradient operator acting $\Delta_a$ is ${\bf \Pi}_a$. In this case, the corresponding expression to eq.(15) is given by 
\begin{eqnarray}
{\rm ln}\biggl(\frac{T}{T_c}\biggr) &+& \int_{\rho_c}^\infty d\rho f(\rho; T) [ 1 - (1 + |\delta N|) \langle I_0(\rho) \rangle ] \nonumber \\ 
&=& - \biggl[ \delta^2 + \biggl(\int_{\rho_c}^\infty d\rho \delta N f(\rho; T_c) \biggr)^2 \biggr]^{1/2}. 
\end{eqnarray}
As this expression independent of $Q$ shows, a complete orbital-limiting is always realized in this case. However, it can be seen that, as far as $\delta \neq 0$, the resulting $H_{c2}(T)$ always lies below that following from eq.(15), implying that, as far as $w_{ss} \neq w_{tt}$, the orbital-limiting with no paramagnetic depairing is not realized. 

%%%%%%%%%%%%%%%%%%%
\begin{figure}[t]
\scalebox{0.35}[0.35]{\includegraphics{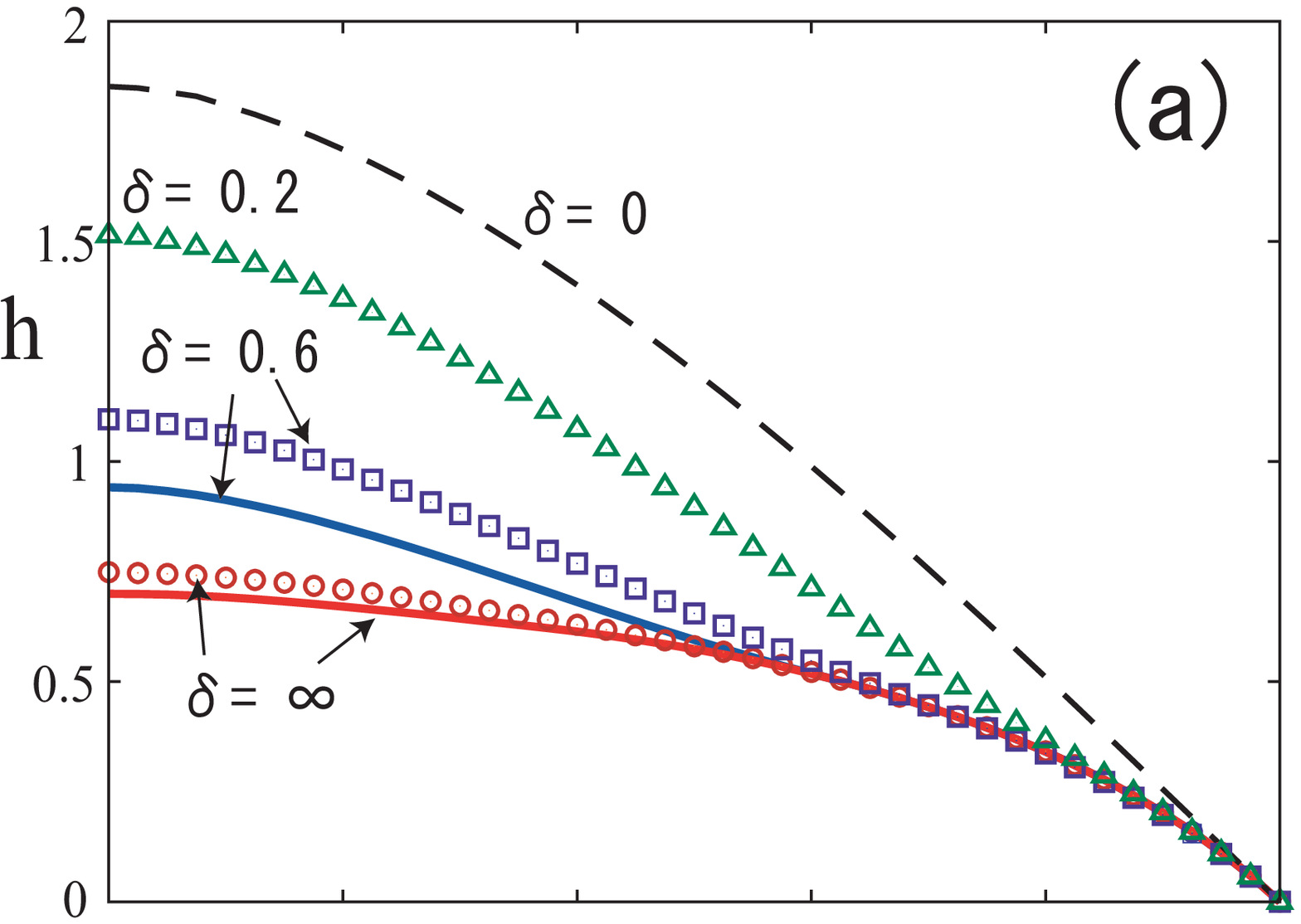}}
\scalebox{0.35}[0.35]{\includegraphics{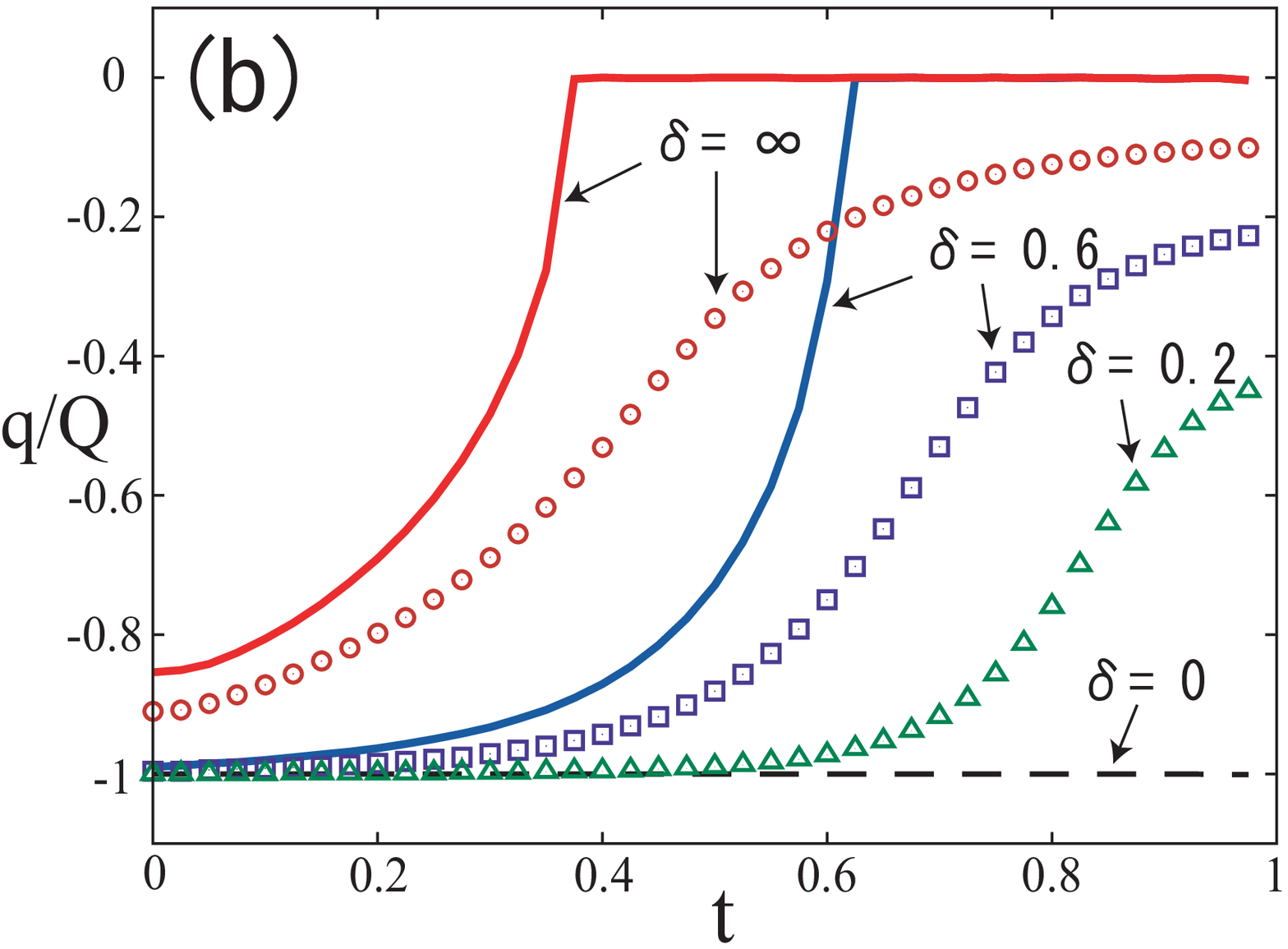}}
\caption{(Color online) (a) Dependences of $H_{c2}(T)$ curves in the cubic case on $\delta$ and $\delta N$, where $h = H/H_{\rm orb}^{({\rm 2D})}(0)$ and $t=T/T_c$. The two solid curves were obtained for $\delta N=0$, while the remaining ones are for $\delta N=-0.1$. The parameters $\mu_{\rm B} H_{\rm orb}^{(2D)}(0)/(2 \pi T_c) = 0.4$ and $\rho_c = (20 \pi T_c)^{-1}$ were used. (b) The corresponding $t$ dependence of $q/Q$ just at $H_{c2}(T)$.} \label{fig.1}
\end{figure}
%%%%%%%%%%%%%%%%%

Now, let us discuss about the $H_{c2}(T)$-curves following from eq.(15). In Fig.1, the $\delta$ dependence of $H_{c2}(T)$ and of the corresponding $q(T)/Q$ just at $H_{c2}(T)$ are shown by setting $\mu_{\rm B} H_{\rm orb}^{({\rm 2D})}(0)/(2 \pi T_c) = 0.4$ and $\delta N = -0.1$, where $H_{\rm orb}^{({\rm 2D})}(T)$ is the orbital-limiting field in 2D case. In the purely singlet or triplet case where $\delta = \infty$, the same magnitude of the energy gap is formed on the two FSs, 
and 
a helical phase modulation parallel to the field at higher temperatures is merely a consequence of a nonvanishing 
$\delta N$ \cite{Kaur}, while the sudden appearance of nonzero $q$ near $t=0.38$ indicates a second order transition into the ordinary FF state \cite{Feigel}. As the lower two curves in Fig.1(a) show, however, a realistic $|\delta N|$-value ($\sim 0.1$) does not lead to a remarkable increase of $H_{c2}$. In contrast, effects of a $\delta$-reduction on $H_{c2}$ and $q$ are more dramatic: Even for $\delta$ of order unity, the $H_{c2}(T)$-enhancement due to the singlet-triplet mixing is much more remarkable than that due to a finite $\delta N$, and the slope of $H_{c2}(T)$ shows a subtle but visible increase below an intermediate temperature upon cooling. As Fig.1(b) shows, this increase of the $H_{c2}$-slope originates from the corresponding increase of $|q|$, i.e., the phase modulation of $\Delta$, upon cooling, and the onset temperature of the $|q|$-growth {\it increases} with decreasing $\delta$. However, this enhancement of the modulation never means that the paramagnetic depairing becomes more effective due to the missing parity. To explain this, we have also examined the corresponding $H_{c2}(T)$ resulting from the LO-type state $\Delta_a = L_z^{-1/2} Z_a({\bf q}) [ e^{{\rm i}qz} \varphi_0(x,y) + e^{-{\rm i}qz} {\tilde \varphi}_0(x,y) ]$ \cite{LO}, where ${\tilde \varphi}_0(x,y)$ may be different from $\varphi_0(x,y)$ but is under the same gauge. In this case, the corresponding expression to eq.(15) for determining $H_{c2}(T)$ is given simply by neglecting the two ${\rm sin}(\rho v_\parallel q) {\rm sin}(\rho v_\parallel Q)$ terms there. Then, by noting that the LO state appears upon cooling at the temperature where the O($q^2$) term of $F_2^{(c)}$ changes its sign, it is easily seen that the onset of the LO state is independent of $\delta$. In contrast, in eq.(15), an additional $q$ dependence appearing as a consequence of a finite $\delta$ (see the r.h.s. of eq.(15)) favors the helical phase modulaton. Thus, the $|q|/Q$-growth enhanced by decreasing $\delta$ in Fig.1 is peculiar to the helical (phase-modulated) state and does not imply an enhancement of the paramagnetic depairing. Rather, the singlet-triplet mixing makes the LO state significanty unfavorable. Further, the curves in Fig.1 also show that an increase of the additional helicity due to a nonzero $\delta N$ \cite{Kaur} becomes more remarkable as the singlet-triplet mixing is increased. 

Through the discussion on $H_{c2}$, we have implicitly assumed the {\it mean field} superconducting transition at $H_{c2}$ to be of second order. To check its validity, let us consider here the corresponding quartic GL term $F^{(c)}_4$. The expression of $F^{(c)}_4$ is briefly explained as follows: Recall that, in the familiar centrosymmetric case with no paramagnetic depairing, $F^{(c)}_4$ within the lowest LL takes the form $\int dx dy |\varphi_0(x,y)|^4 \langle \int \Pi d\rho_j F({\rho_j}; {\hat p}) \rangle$ and is positive. Here, ${\hat p}$ denotes a unit vector parallel to a momentum on each FS. In the weak coupling approximation, $F_4^{(c)}$ in the present cubic case is simply a sum of contributions from the two FSs and is given by  
\begin{eqnarray}
 F_4^{(c)} &=& \int dx dy |\varphi_0(x,y)|^4 \sum_{a=1,2} N_a Z_a^4 \langle \int \Pi d\rho_j \nonumber \\
 &\times& {\rm cos}(\sum_j \rho_j v_\parallel(q - (-1)^a Q)) F({\rho_j}; {\hat p}) \rangle,
\end{eqnarray}
which is clearly positive since $q < 0$ when $\delta N < 0$. Actually, we have verified that $F_4^{(c)}$ is always positive within the calculations performed by us. This fact implies that there is no occasion that the second order transition at $H_{c2}$ assumed above is preempted by a discontinuous transition \cite{AI} in the present cubic noncentrosymmetric case. 

\section{Transverse magnetization in cubic case} 

In addition to the $H_{c2}$-line, an observable measure of the phase modulation of the vortex state in the cubic noncentrosymmetric case will be needed to verify its presence in real systems. Recently, the presence of a nonvanishing component ${\bf m}_\perp$ {\it perpendicular} to ${\bf H}$ of the local magnetization ${\bf m}$ in the ordinary FFLO vortex lattice has been stressed within the gradient expansion approach \cite{Mineev} valid for sufficiently large Maki parameters. In the phase-modulated FF state, this transverse magnetization ${\bf m}_\perp$ occurs from a nonvanishing periodic component $j_\parallel({\bf r})$ of the current {\it parallel} to ${\bf H}$ \cite{Mineev,Yuan}. In this section, our result of a quantity corresponding to ${\bf m}_\perp$ will be shown. This result will clarify the implication of the $H_{c2}$-enhancement due to the phase modulation in sec.II further. 

The local supercurrent ${\bf j}$ is given by 
\begin{eqnarray}
&{\bf j}_\mu& = -\frac{\delta {\cal F}^{(c)}_2}{\delta {\bf A}}\biggr|_{\delta {\bf A}=0} = 8 e \int_0^\infty d\rho f(\rho; T) \sum_{a=1,2} \langle {\rm i} \rho {\bf v}_\mu \varphi_0^*(x,y) \nonumber \\  &\times& \exp({\rm i} \rho ({\bf v}_\perp\cdot{\bf \Pi} + v_\parallel (q - (-1)^a Q)))  \varphi_0(x,y) \rangle N_a Z_a^4.
\end{eqnarray}
Below, each component of ${\bf j}$ will be expressed as
\begin{eqnarray}
{\bf j}_\perp &=& \sum_{{\bf K} \neq 0} {\bf j}_\perp({\bf K}) F_{\bf K} e^{{\rm i}{\bf K}\cdot{\bf r}}, \nonumber \\
j_\parallel &=& \sum_{{\bf K} \neq 0} j_\parallel({\bf K}) F_{\bf K} e^{{\rm i}{\bf K}\cdot{\bf r}},
\end{eqnarray}
where $F_{\bf K}$ is the Fourier transform of $|\varphi_0(x,y)|^2$, ${\bf K}$ is the reciprocal lattice vector of the vortex lattice, and 
\begin{eqnarray} 
{\bf j}_\perp({\bf K}) &=& {\rm i} 8 e \int_0^\infty d\rho \rho f(\rho; T) \sum_{a=1,2} N_a Z_a^2 \langle {\bf v}_\perp {\rm cos}(\rho v_\parallel (q \nonumber \\ 
&-& (-1)^a Q)) I_0(\rho) \exp(-\rho({\bf v} \times {\bf K})_z/2) {\rm cos}(\rho {\bf v}\cdot{\bf K}/2) \rangle, \nonumber \\
j_\parallel({\bf K}) &=& 8 e \int_0^\infty d\rho \rho f(\rho; T) \sum_{a=1,2} N_a Z_a^2 \langle (-v_\parallel) I_0(\rho) \nonumber \\
&\times& {\rm sin}(\rho v_\parallel (q - (-1)^a Q))  \exp(-\rho({\bf v} \times {\bf K})_z/2) \nonumber \\
&\times& {\rm cos}(\rho {\bf v}\cdot{\bf K}/2) \rangle. 
\end{eqnarray}
We note that, in centrosymmetric superconductors occuring on a single Fermi surface where $q=0$, and $N_a = N(0)$, ${\bf j}_\perp$ in eq.(23) reduces to the corresponding expression ${\bf j}^{(0)}$ of the familiar Abrikosov lattice in the $n=0$ LL in the orbital-limited case with $Q=0$. Then, the local magnetization vector is given by $({\bf h} - {\bf H})/(4 \pi)$ and is related to the current through the Maxwell equation $\nabla \times {\bf h} = 4 \pi {\bf j}$. Below, we will focus on estimating not ${\bf m}_\perp$ itself but rather the normalized $j_\parallel$ which will be defined here as $J = j_\parallel(0, K_{0, y})/|j^{(0)}_x(0, K_{0,y})|$, where $K_{0,y}=\pi^{1/2}/(3^{1/4} r_H)$ is the magnitude of the smallest reciprocal lattice vectors. The magnitude and sign of ${\bf m}_\perp$ is determined by $J$. 
%%%%%%%%%%%%%%%%%%%
\begin{figure}[t]
\scalebox{0.35}[0.35]{\includegraphics{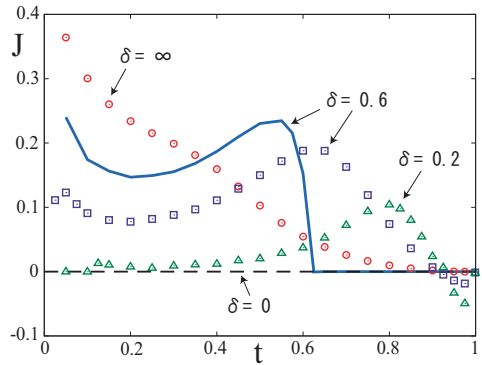}}
\caption{(Color online) Calculated curves of $J = j_\parallel(0,K_{0,y})/|j^{(0)}_x(0,K_{0,y})|$ at $H_{c2}(T)$ obtained in terms of the data in Fig.1. The same symbols and the type of the lines as in Fig.1 are used here.} \label{fig.2}
\end{figure}
%%%%%%%%%%%%%%%%%

In Fig.2, our calculation results of the normalized quantity $J$ at $H_{c2}(T)$-curves are shown. They are obtained by combining the data in Fig.1 into the above expressions. The curves in Fig.2 should be fingerprints of features peculiar to the vortex lattice occuring at least near $H_{c2}$ in the cubic noncentrosymmetric case. As is found by comparing Fig.2 with Fig.1, a growth of $|q|$ induced by the mixing of the singlet and triplet pairings results in a reduction of $|{\bf m}_\perp|$, i.e., of the paramagnetic depairing, while this reduction of the paramagnetic effect is safely negligible in $\delta=\infty$ case with no mixing between the pairing channels, and $|{\bf m}_\perp|$ monotonously increases upon cooling reflecting an enhancement of the paramagnetic effect upon cooling. Thus, the nonmonotonous $t$-dependence of $J$ seen in $\delta=0.2$ and $0.6$ cases at lower temperatures is a consequence of the competition between an enhancement of the paramagnetic depairing upon cooling and its effective reduction due to a growth of $|q|$. 

\section{Depairing field of quasi 2D Rashba superconductors} 

Next, let us briefly explain the corresponding results for a Rashba superconductor with the basal plane perpendicular to ${\hat z}$ in ${\bf H} \parallel {\hat y}$ (i.e., a parallel field configuration). In this case, ${\hat {\bf g}}_{\bf k} = ({\bf k} \times {\hat z})/k_F$, and the unitary matrix $U({\bf k})$ is replaced by $(1 + {\rm i}({\rm sin}\phi_{\bf k} \sigma_y - {\rm cos}\phi_{\bf k} \sigma_x))/\sqrt{2}$. Then, $\Psi^{(s)}_{\bf p}$ in eq.(4) needs to be replaced by ${\rm i} \, \Psi^{(s)}_{\bf p}$. consequently, the off-diagonal element $w_{st}$ of the interaction matrix appears only in the "intraband" terms of the GL free energy (see below). Further, following the purely singlet (or purely triplet) case \cite{ymatsu}, we use a cylindrical FS extending and corrugating along ${\hat z}$. Then, the factor $|{\hat {\bf g}}_{\bf k}|$ in $\Psi^{(t)}_{\bf p}$ of eq.(4) may be replaced by unity. The quadratic term $F^{({\rm R})}_2$ of the resulting GL free energy, corresponding to eq.(8) in the cubic case, reduces to eq.(2) in Ref.\cite{ymatsu} in $w_{tt}$, $w_{st} \to 0$ limit. By noting that, in $F^{({\rm R})}_2$ with ${\bf A} = H z {\hat x}$, the gauge-invariant operator $- {\rm i}\nabla + (r_H^{-2} z + (-)^a Q) {\hat x}$ acts on $\Delta_a \equiv (\Delta_s + (-)^{a+1} \Delta_t)/\sqrt{2}$ ($a=1$, $2$), it is convenient to express $\Delta_a$ in terms of LLs {\it dependent on} the two FSs in the manner $
\Delta_a = \sum_{n \geq 0} Y_{a,n} \, \, \varphi_n(z + (-)^a Q  r_H^2 
, \, \, x)$. Then, we have 
\begin{widetext}
\begin{eqnarray}
\frac{F^{({\rm R})}_2}{2V} &=& \sum_{n,a} \biggl( \frac{(w^{-1})_{ss} + (w^{-1})_{tt}}{2} + (-)^{a+1} (w^{-1})_{st} 
- \int_{\rho_c}^\infty d\rho \frac{4 \pi T N_a}{{\rm sinh}(2 \pi T \rho)} \langle I_n(\rho) \rangle \, \biggr)|Y_{a,n}|^2 \nonumber \\
&+& ( \, (w^{-1})_{ss} 
- (w^{-1})_{tt} \, ) \sum_{n_1,n_2} \frac{W_{n_1,n_2}(Q)}{2} \, ( \, Y_{2,n_1}^* Y_{1,n_2} 
+ {\rm c.c.} ),
\end{eqnarray} 
\end{widetext}
where 
\begin{widetext}
\begin{eqnarray}
W_{n,m}(Q) &=& \int dz dx \, \varphi^*_m(z,x) \, \varphi_n(z + 2Qr_H^2, x) 
= \exp(-Q^2 r_H^2)  \sum_{l=0}^{{\rm min}(m,n)} \frac{(-1)^{m-l} \sqrt{n! m!}}{(n-l)! (m-l)! \, l!} (\sqrt{2}Q r_H)^{n+m-2l}, 
\end{eqnarray}
\end{widetext}
and the zero field $T_c$ is determined by 
\begin{widetext}
\begin{eqnarray}
\frac{(w^{-1})_{ss} + (w^{-1})_{tt}}{2(N_1+N_2)} &=& \int_{\rho_c}^\infty d\rho \, f(\rho; T_c) + \biggl[ \biggl(\frac{(w^{-1})_{ss} - (w^{-1})_{tt}}{2(N_1+N_2)} \biggr)^2 
+ \biggl(\int_{\rho_c}^\infty d\rho \, \delta N \, f(\rho; T_c) - \frac{(w^{-1})_{st}}{(N_1+N_2)} \biggr)^2 \biggr]^{1/2}.
\end{eqnarray}
\end{widetext}
%%%%%%%%%%%%%%%%%%%
\begin{figure}[t]
\scalebox{0.35}[0.35]{\includegraphics{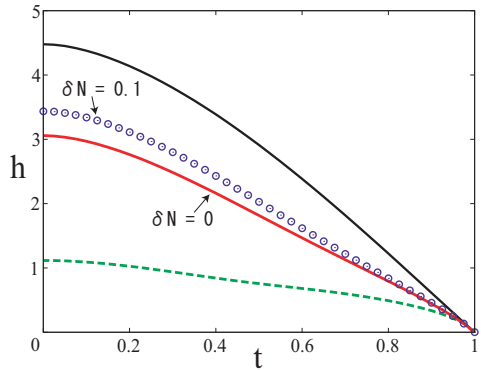}}
\caption{(Color online) (a) Examples of calculated $H_{c2}(T)$ curves in the Rashba case with $\mu_{\rm B} H_{\rm orb}^{(2D)}(0)/(2 \pi T_c) = 0.4$ in ${\bf H} \perp c$. The used parameters ($|(w^{-1})_{tt} - (w^{-1})_{ss}|/(N_1+N_2)$, $\delta N$) are ($0$, $0$) (black solid curve), ($0.4$, $0$) (red solid curve), ($0.4$, $-0.1$) (open circles), and ($\infty$, $-0.1$) (dashed curve). For simplicity, we have put $w_{st}=0$. } \label{fig.3}
\end{figure}
%%%%%%%%%%%%%%%%%%

In this Rashba case, a direct numerical evaluation is needed to examine $H_{c2}$. Nevertheless, the main result is already found in eq.(25) in the specific $w_{ss}=w_{tt}$ case where the "interband" term is absent: In eq.(25), the paramagnetic depairing appearing only through the $Q$-dependences in the interband terms is completely quenched {\it irrespective of} the $w_{st}$-value in the $w_{ss} \to w_{tt}$ limit where the pairing occurs just on the FS $1$ with a larger density of states. Then, the vortex state close to the resulting $H_{c2}$ is described primarily by the Abrikosov triangular lattice solution $\Delta_1$ with {\it no} additional modulation of the order parameter. 
Even in this Rashba case, a discontinuous normal to superconducting transition \cite{AI} is expected not to occur, because, as seen above, the paramagnetic effect is lost in the limit where the pairing interactions in the singlet and triplet channels occur with a comparable weight, while the absence of the temperature region with such a discontinuous transition was verified in the opposite limit with just a single pairing state \cite{ymatsu}. Therefore, an exotic sequence of structural transitions between different vortex lattices \cite{ymatsu} may not appear in systems with a small $|(w^{-1})_{ss} - (w^{-1})_{tt}|$ close to the orbital-limiting. Examples of $H_{c2}(T)$ curves obtained by directly 
diagonalizing eq.(25) in terms of the lower 
six ($0 \leq n \leq 5$) LLs 
are shown in Fig.3. It seems that the $H_{c2}(T)$ curves at least in $t < 0.6$ are not dependent much on whether still higher LLs are included or not. It is found that the $w_{ss} - w_{tt}$ dependence of $H_{c2}(T)$ curves is qualitatively similar to the $\delta$ dependence in the cubic case, and that a slight but visible slope change of $H_{c2}(T)$ near $T \simeq 0.65 T_c$ occurs for a smaller $|\delta N|$. Further, in the case with $\delta=\infty$, i.e., with just a single pairing channel, a clear slope change of $H_{c2}(T)$ is detectable near $t=0.5$ and is closely related to a structural transition between vortex lattices \cite{ymatsu}. This result will be relevant to a similar behavior seen in $H_{c2}$ data of CeRhSi$_3$ in ${\bf H} \perp c$ \cite{Kimura}. Details of the corresponding vortex lattice structures will be reported elsewhere. 

\section{Summary and Discussion} 

In the preceding sections, we have shown that, irrespective of the form of the broken inversion symmetry, the mixing and the {\it field-induced} coupling between coexisting singlet and triplet pairing states significantly suppress the paramagnetic depairing effect and lead to an enhancement of $H_{c2}$. In the noncentrosymmetric systems of the cubic or Rashba type, the paramagnetic effect enters as an additional gauge field in the gradient ${\bf \Pi}= -{\rm i}\nabla + 2 e {\bf A}$ acting on the order parameter $\Delta$ in the manner dependent on the FSs; ${\bf \Pi} + {\bf Q}$ for one FS and ${\bf \Pi} - {\bf Q}$ for the other. If either of the two splitted FSs is irrelevant to superconductivity, ${\bf Q}$ is trivially gauged away, and the paramagnetic term plays no roles of a pair breaking. In contrast, when both of the two FSs contribute to superconductivity, the gauge fields $\pm {\bf Q}$ {\it frustrate} with each other and are not cancelled by a gauge transformation so that the paramagnetic depairing effectively works. In these noncentrosymmetric systems, either of the two FSs may become irrelevant to superconductivity 
when both of a singlet and a triplet pairing channels have attractive interactions of the same order in magnitude, and then, the orbital-limited $H_{c2}$ is realized. 

It will be valuable here to explain relations of the present work with other previous ones addressing noncentrosymmetric superconductors in nonzero fields. In Ref.\cite{Frigeri}, the coexistence of a singlet and a triplet pairing channels was taken into account in the Pauli limit where ${\rm i} \, {\bf \Pi}$ (see the preceding paragraph) is replaced by the gradient $\nabla$ so that the vortices are ignored. Further, any modulation of $\Delta$, i.e., any contribution of the gradient $\nabla \Delta$, was ignored there, and consequently, the strength of coupling between the two pairing states was measured only by $|\delta N| \simeq \zeta/E_F$ as in zero field case. However, this result is invalidated once a modulation of $\Delta$ is taken into account. Some results of a treatment in the Pauli limit taking account of contributions of $\nabla \Delta$ were commented on in Ref.\cite{Agter} by focusing on the Rashba case. It seems that a divergence of $H_{c2}$ for a cylindrical FS, noted there \cite{Agter}, at an intermediate temperature in the case with both of a singlet and a triplet pairing channels corresponds to an orbital-limited situation found here in sec.IV by taking account of the vortices. However, any physical implication of the $H_{c2}$-divergence and details of calculations leading to such results were not explained there \cite{Agter}. The crucial point is that the coupling, induced by the magnetic field and a $\Delta$-modulation, is present between the two pairing channels even in $\delta N \to 0$ limit. 

Finally, we discuss about relevance of the present results to real systems. As noted in sec.I, two Rashba superconductors, CeRhSi$_3$ \cite{Kimura} and CeIrSi$_3$ \cite{Settai}, show a strong paramagnetic effect in a parallel field, and their in-plane $H_{c2}(0)$ values (in ${\bf H} \perp c$) are significantly suppressed compared with that in ${\bf H} \parallel c$. In contrast, the $H_{c2}$ lines in CePt$_3$Si and LaIrSi$_3$ are nearly isotropic and show no sign of the paramagnetic depairing even in the parallel fields. In particular, CePt$_3$Si has a large effective mass of the normal quasiparticles, and, in fact, the Pauli-limiting field $H_P(0)$ was estimated to be much lower than $H_{c2}(0)$ in all configurations \cite{Yasuda}. It will be reasonable to, according to the present results, attribute the apparent absence of paramagnetic depairing in this material under a parallel field to a mixing of two pairing channels 
with a {\it comparable} weight. It is quantitatively insufficient to regard the nearly isotropic $H_{c2}$ \cite{Yasuda} as a consequence of a finite $\delta N$ \cite{Kaur}. The present view on the pairing state of CePt$_3$Si, following from studies of the $H$-$T$ phase diagram, is consistent with a recent proposal \cite{Frigeri2} based on microscopic properties. 
However, it is unclear at present whether the neglect in the present work of an antiferromagnetic order \cite{Yasuda,Bauer} existing in CePt$_3$Si is justified or not. According to calculations in Ref.\cite{Yanase} where the vortices are ignored, the presence of an antiferromagnetic order may lead to a significant deviation of the $d$-vector from ${\hat {\bf g}}_{\bf k}$ and result in some reduction of the paramagnetic depairing effect. 
%In contrast, an apparent absence of paramagnetic depairing was found in LaPt$_3%$Si with no magnetic order. However, it is possible that, judging from the NMR %data \cite{Kitaoka}, the component of a $s$-wave pairing is more or less domina%nt in this material, and details 
%of this material should be clarified further. 

At present, the best candidate for applying the present results in the cubic case will be the familily of Li$_2$(Pd$_{3-x}$Pt$_x$)B \cite{Yuan,Hirata}. According to a recent study of $H_{c2}$-curves of these materials \cite{Hirata}, however, they seem to be well explained in the weak coupling approximation with no paramagnetic effect and in clean limit. Since Li$_2$Pd$_3$B is believed to be in the purely $s$-wave pairing, this fact may suggest an extremely weak paramagnetic effect in these materials. Nevertheless, the transverse component of the magnetization, stressed in sec.III, might be measurable, and its experimental search is hoped. 

\begin{acknowledgements}
We are grateful to K. Hirata for providing us with a copy 
of Ref.\cite{Hirata}. 
\end{acknowledgements}

\end{document}